\documentclass[a4paper,12pt]{iopart}

\usepackage{amssymb}
\usepackage{mathrsfs}
\usepackage{lineno}
\usepackage{subfigure}
\modulolinenumbers[5]
\usepackage{graphicx}
\usepackage{latexsym}
\usepackage{mathdots}
\usepackage{bm}
\usepackage{epstopdf}
\usepackage{tabularx} 
\usepackage{multirow}
\usepackage{float}
\usepackage{array}
\usepackage{cite}
\usepackage{siunitx}
\usepackage{xcolor}
\usepackage{pdflscape}

\newcommand{\w}{\omega}

\begin{document}

\title[]{Small Angle and Non-Monotonic Behavior of the Thermal Conductivity in 
Twisted Bilayer Graphene}

\author{Chenyang Li \& Roger K. Lake}

\address{Department of Electrical and Computer Engineering, University of California, Riverside, California 92507, USA}
\ead{rlake@ece.ucr.edu}
\vspace{10pt}
\begin{indented}
\item[]November 2019
\end{indented}

\begin{abstract}
Nothing is known about the thermal conductivity in twisted bilayer graphene (TBG) at small twist angles,
and how it approaches its aligned value as the twist angle approaches $0^\circ$.
To provide insight into these questions,
we perform large scale non-equilibrium molecular dynamics calculations on commensurate
TBG structures with angles down to $1.87^\circ$.
The results show a smooth, non-monotonic behavior of the thermal conductivity with respect to the
commensurate lattice constant. 
As the commensurate lattice constant increases, the thermal conductivity initially decreases 
by 50\% and then returns to 90\% of its aligned value as the angle is reduced to $1.89^\circ$.
These same qualitative trends are followed by the trends in the shear elastic constant,
the wrinkling intensity, and the out-of-plane ZA$_2$ phonon frequency.
The picture that emerges of the physical mechanism governing the thermal conductivity 
is that misorientation reduces the shear elastic constant; the reduced shear elastic constant enables
greater wrinkling; and the greater wrinkling reduces the thermal conductivity.
The small-angle behavior of the thermal conductivity raises the question of how do response functions
approach their aligned values as the twist angle approaches $0^\circ$. 
Is the approach gradual, discontinuous, or a combination of the two?
\end{abstract}

%
%
%
%
%

\section{Introduction}
The effect of layer misorientation on the electronic structure and the electrical conductance 
of bilayer graphene (BLG) and multi-layer graphene has received much 
attention \cite{mele2010commensuration,shallcross2010electronic,Flat_Bands_TGB_Morell_PRB10,MacDonald_PNAS11,bistritzer2010transport,Habib_Graphite_APL13,koren2016coherent,Shepard_NL16,Bernevig_All_Angles_Topo_PRL19}, and 
interest was recently renewed by the experimental discovery of superconductivity at 
certain low misorientation angles \cite{cao2018unconventional,Tuning_SC_TBG_CDean_Sc19}
where the electronic bands become flat at the Fermi level \cite{Flat_Bands_TGB_Morell_PRB10,MacDonald_PNAS11}.
Experimentally, the effect of interlayer rotation on the phonon spectrum
has been probed extensively with Raman 
spectroscopy \cite{carozo2011raman, havener2012angle, kim2012raman, he2013observation, campos2013raman, lui2012observation, coh2013theory, righi2013resonance,ramnani2017raman,eliel2018intralayer}.
Theoretical research on the phonon properties of twisted bilayer graphene (TBG) finds that the phonon frequencies, 
density of phonon modes, phonon velocities, and specific heats of the low
frequency phonon branches vary little with the interlayer rotation 
angle \cite{cocemasov2013phonons,ramnani2017raman,cocemasov2015engineering,nika2014specific,li2018commensurate}.
Bringing these two different lines of research together, recent theory proposes a phonon driven mechanism
for the superconductivity \cite{Bernevig_Phonon_Driven_TBG_Superconductor}.
The effect of misorientation on the in-plane thermal conductivity of TBG has received less attention
\cite{li2014thermal,mao2016atomistic,minthermal2017,li2018commensurate,nie2019interlayer}.
The one experimental study on the in-plane thermal conductivity of TBG carried out
opto-thermal measurements \cite{ghosh2008extremely} 
on one TBG sample with a twist angle of $\sim 32^\circ$, and found that interlayer 
misorientation reduced the
in-plane thermal conductivity by up to 50\% \cite{li2014thermal}.

Standard expressions for the thermal conductivity based on the phonon Boltzmann transport equation
show that the lattice thermal conductivity depends on the phonon velocities, frequencies, and lifetimes.
Since misorientation has little effect on the phonon velocities and frequencies, 
it was proposed that that the zone-folding that occurs in TBG
opens up new channels for phonon scattering that are unavailable in unrotated BLG \cite{li2014thermal}.
As a consequence, the phonon lifetimes are reduced, which results in a reduction in the thermal conductivity.
A recent theoretical study of the of the lattice thermal conductivity of TBG with three rotation angles corresponding
to the three smallest commensurate unit cells, $21.78^\circ$, $32.17^\circ$, and $13.17^\circ$, 
found that the thermal conductivity decreased approximately linearly as the commensurate lattice constant 
increased \cite{li2018commensurate}.
The scaling of the thermal conductivity with the lattice constant rather than the angle
was consistent with the hypothesis that the decreased thermal conductivity in TBG resulted 
from increased scattering allowed by the large zone-folding in the reduced Brillouin zones \cite{li2014thermal}.

However, the three angles considered in \cite{li2018commensurate} give only a small picture within the total range
of possible misorientation angles.
What happens at smaller misorientation angles of $10^\circ$ or less is still an open question.
It seems reasonable to expect that as the rotation angle is reduced towards zero, 
the thermal conductivity might return to its aligned
value in some smooth manner even though the commensurate lattice constant becomes very large. 
If this expectation were true, then there would be a minimum in the thermal conductivity as a function
of the commensurate lattice constant.
Such a non-monotonic dependence of the thermal conductivity on the commensurate lattice constant would suggest
that physical mechanisms other than increased scattering allowed by reduced Brillouin zones
play a role in governing the thermal conductivity.
To investigate the mechanisms that govern the thermal conductivity in TBG,
and to provide insight into the physical mechanisms that give rise to the angle
and lattice constant dependence of the thermal transport, 
we perform large-scale non-equilibrium molecular dynamics calculations of the thermal conductivity 
of TBG for commensurate twist angles down to $1.89^\circ$, we calculate the elastic constants
and the phonon spectra for the misoriented structures, and we compare the results to those from
other theoretical and experimental works.
\begin{figure}[H]
\centering
\vspace{0.5cm}
\includegraphics[width=5.0in]{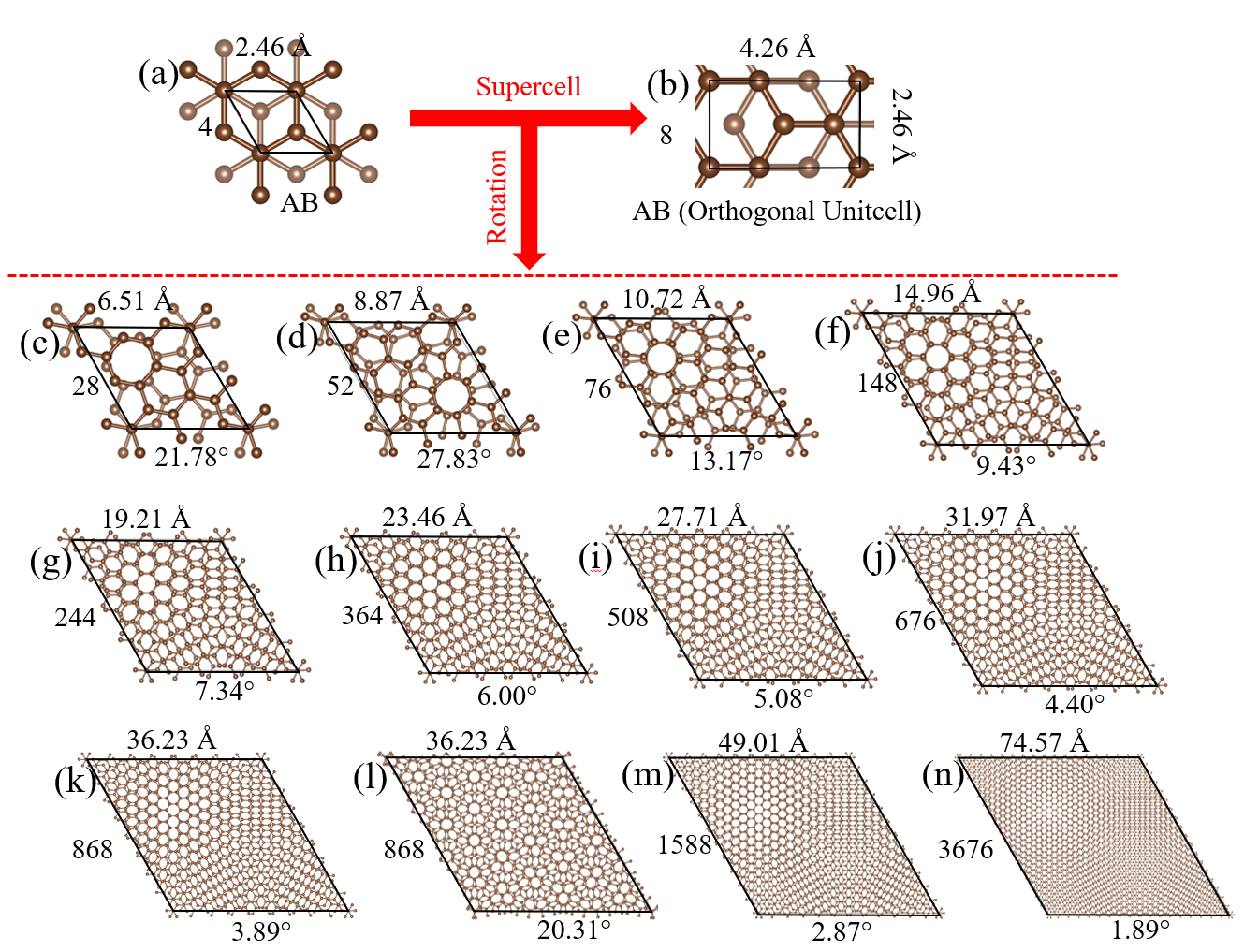} 
\caption{
(a) and (c-n) 
The primitive cells of AB-BLG and commensurate TBG. 
For each primitive cell, 
the rotation angle is given along the bottom edge,
the number of atoms are shown along the left edge, and
the commensurate lattice constant is given along the top edge.
(b) A rectangular unit cell created and then repeated for constructing the 
long ribbons for the thermal transport calculations.
}
\label{fig:models} 
\end{figure} 

\section{Methods}
\label{sec:Methods}
The approach used to create the commensurate unit cells and to model their thermal conductivities 
was previously described
in detail \cite{li2018commensurate}, and only a brief description of the most important points is provided here.
A total of 13 different commensurate rotation angles are considered with their commensurate primitive cells
shown in Fig. \ref{fig:models}.
For each structure, the rotation angle, lattice constant, and number of atoms in the commensurate primitive cell are shown. 
The sizes of the commensurate primitive cells quickly increase as the rotation angle decreases, 
and at the smallest angle of 1.89$^\circ$, the primitive cell contains 3676 atoms.
All of the angles chosen fall along the curve of minimum commensurate primitive cells shown in Fig. 2 of Shallcross
et al. \cite{shallcross2010electronic}, except for the one angle of $20.31^\circ$.
The primitive commensurate cells for $20.31^\circ$ and $3.89^\circ$ have the same primitive commensurate cell
lattice constants, even though
their moir\'{e} patterns look very different.
The misoriented primitive cells for all of angles that fall along the curve of minimum commensurate primitive cells appear
to smoothly transition from a region of AB stacking to a region of AA stacking.
For the $20.31^\circ$ structure, there are many such transitions within the primitive cell. 
This angle is included to test whether the physics governing the thermal conductivity is determined
by the rotation angle or the size of the primitive commensurate cell.
If the physics is governed by the rotation angle, then the thermal conductivities for misorientations
of $3.89^\circ$ and $20.31^\circ$ should be very different.
If the physics is governed by the size of the commensurate primitive cell, then the thermal conductivities should be
the same.

Calculations of the phonon dispersions and thermal conductivities are performed using molecular dynamics (MD)
and non-equilibrium molecular dynamics (NEMD) \cite{muller1997simple} as implemented in LAMMPS \cite{plimpton2007lammps}.
Detailed benchmarking of various interatomic potentials has been reported for graphene \cite{koukaras2015phonon},
but there are no equivalent benchmarking studies for bilayer graphene.
The common intralayer potentials include Tersoff \cite{tersoff1989modeling, lindsay2010optimized}, 
Brenner \cite{Brenner_PRB90}, the reactive empirical bond order potential 
(REBO) \cite{stuart2000reactive, brenner2002second},
and the long-range bond-order potential for carbon (LCBOP) \cite{los2003intrinsic}.
All of these potentials belong to empirical bond order potentials (EBOPs) \cite{abell1985empirical} 
and treat electronic binding as effective pairs.
For bilayer graphene and misoriented bilayer graphene, the long range interlayer potential is critical. 
To model this, a long range interlayer potential is added to the above intralayer potentials,
and it generally takes the form of a Lennard-Jones (LJ) potential \cite{stuart2000reactive}. 
Most recently a new interlayer potential, dihedral-angle-corrected registry-dependent interlayer potential
(DRIP), was created specifically for misoriented multilayer graphene \cite{wen2018dihedral}.
Calculations presented here used REBO for the intralayer potential, 
which is the most recent extension originating from the Tersoff
potential, with the two types of interlayer potentials, LJ, as implemented in
the adaptive intermolecular REBO (AIREBO) potential, and DRIP.
All calculations are performed with AIREBO, and the main result, the
trend in the thermal conductivity, is verified with REBO+DRIP.

The thermal conductivity is calculated using non-equilibrium molecular dynamics (NEMD)
implemented in LAMMPS \cite{plimpton2007lammps}, in which
a constant small heat flux is applied across the simulation domain and the gradient of the average temperature
directly gives the thermal conductivity.
The average temperature for all calculations is $T=300$ K.
For calculation of the thermal conductivity using this direct approach, the primitive cells 
shown in Fig. \ref{fig:models} are expanded into rectangular cells, 
and the rectangular cells are then repeated in both length and width to form long ribbons for the simulation domain.
Periodic boundary conditions are used in the width direction so that there are no edges and no edge effects.
When we refer to the ``width'' of the ribbon, we are referring to the width of the central ribbon to which we apply
periodic boundary conditions.

\begin{figure}[H]
\centering
\vspace{0.5cm}
\includegraphics[width=5.0in]{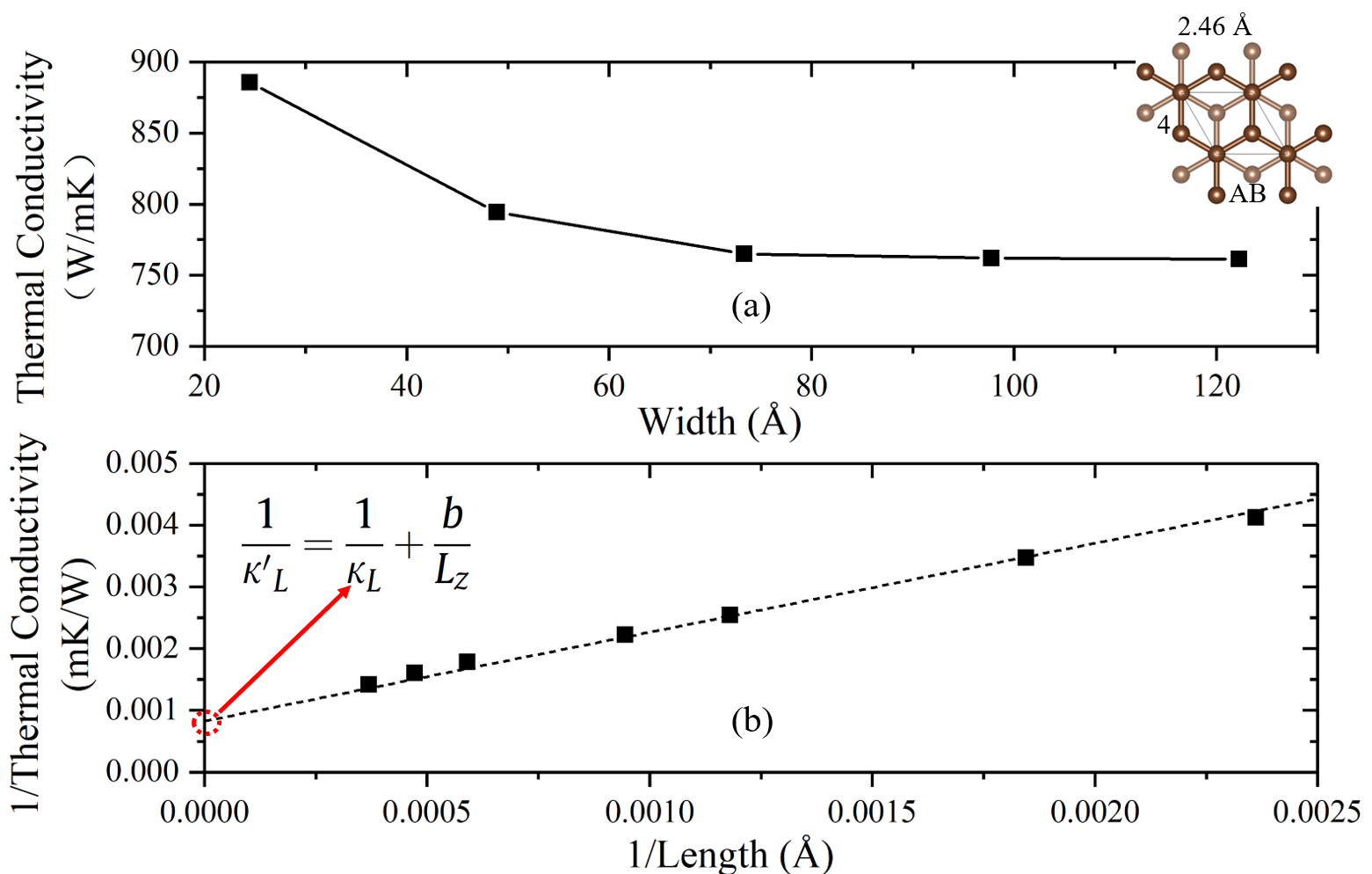} 
\caption{Lattice thermal conductivity of AB-BLG plotted as a function of (a) width and (b) inverse length. 
The length of the ribbon in (a) is $1.06$ $\mu$m and the width of the ribbon in (b)
is $7.87$ nm.
}
\label{fig:size effect} 
\end{figure}

Finite width and finite length effects are both present, and they are addressed using the following approaches.
The thermal conductivity of ribbons of increasing width are simulated until the thermal conductivity $(\kappa_l)$ 
becomes independent of the width.
The width at which this occurs is $\sim 70$ {\AA} as shown in Fig. \ref{fig:size effect}(a). 
All of the simulated structures have a width greater than $70$ {\AA}. 
The widths slightly vary, since the ribbons must be constructed from integer multiples of the primitive
cells shown in Fig. \ref{fig:models}.
%
%

To address the finite length effect, for each angle in Fig. \ref{fig:models}, multiple ribbons are
constructed of increasing length $L$ ranging from 20 nm to 12.9 $\mu$m. 
The largest ribbon contains $8,359,232$ atoms.
The inverse of the calculated thermal conductivity $\frac{1}{\kappa_L^\prime}$ for each length is plotted versus $1/L$
and fit to the line 
$\frac{1}{\kappa_L^\prime} = \frac{1}{\kappa_L} + \frac{b}{L}$ \cite{huang2006evaluation}. 
As shown in Fig. \ref{fig:size effect}, the dependence of $\frac{1}{\kappa_L^\prime}$ on $\frac{1}{L}$ is linear.
The intercept at $\frac{1}{L}=0$ gives the converged value of $\kappa_L$ as $L \rightarrow \infty$. 
Phonon dispersions are calculated using the Fix-phonon package of the LAMMPS code \cite{kong2011phonon}.
In this approach the dynamical matrix is constructed directly from the time averaged 
displacement-displacement correlation function
evaluated during the molecular dynamics simulations. 
The dynamical matrix constructed in this manner is temperature dependent, and all simulations
are performed at a temperature of $T=300$ K.
To avoid negative phonon frequencies near $\Gamma$, 
we use $25 \times 25\times 1$ supercells for all structures.
The resulting supercell sizes range from 17,500 atoms (AB) to 2,229,750 atoms ($1.89^\circ$).
The total number of iterations to enforce the acoustic sum rule is set at 50.
All settings are the same as those used in the NEMD simulations.
Velocities of the three acoustic branches are determined by evaluating the derivatives,
$v = \left. \frac{\partial \omega}{\partial q} \right|_{q \rightarrow 0}$.

The elastic constants are obtained by introducing small deformations of 
the crystal cell around the equilibrium configuration and solving
\begin{equation}
\frac{\delta{E}}{V_0}=\frac{1}{2}\sum_{ij,kl}C_{ij,kl}\epsilon_{ij}\epsilon_{kl},
\quad \textrm{and} \quad
\epsilon_{ij} = \frac{1}{2}(\frac{\partial{\delta{x_i}}}{\partial{x_j}}+\frac{\partial{\delta{x_j}}}{\partial{x_i}})
\label{eq:deformation}
\end{equation}
where $i$, $j$, $k$, $l$ are deformation directions in three dimensions, 
$C_{ij,kl}$ is the elastic constant, 
$V_0$ is the equilibrium volume of the relaxed structure, 
and $\delta{E}$ is the potential difference recorded at each timestep.
The Voigt form for the elastic constants will be used in the results and discussion; 
for example $C_{44}$ is the Voigt notation for $C_{23,23}$.

The out-of-plane wrinkling intensity is quantified with the unitless metric 
$\gamma = (\eta_{A}/\eta_{\lambda})\times 100\%$,
where $\eta_{A}$ is the mean wrinkling amplitude and $\eta_{\lambda}$ is the mean wrinkling 
wavelength \cite{wang2014anisotropic}.
$\eta_{A}$ is obtained by the averaging the standard deviation of out-of-plane coordinates 
of every atom in each layer, 
$\eta_{A} = \frac{1}{2}\sum_{l=1}^{2}\sqrt{\frac{1}{N}\sum_{i=1}^{N}(z_{i,l}-\bar{z}_l)^2}$,
where $l$ denotes the layer number, $N$ is the total number of atoms, 
$z_{i,l}$ is the out-of-plane coordinate of atom $i$ in layer $l$, and 
$\bar{z}_l$ is the average out-of-plane coordinate of layer $l$. 
The wrinkling wavelength $\eta_{\lambda}$ is determined from the 
Fourier transform of $z_i$ along the heat transfer direction, 
$\frac{1}{N} \sum_{j,l} (z_{j,l} - \bar{z}_l) e^{ikx_{j,l}}$, where $x_{j,l}$ is the $x$ coordinate of atom $j$
in layer $l$.

\section{Results}
\label{sec:Results}

\begin{figure}[H]
\centering
\vspace{0.5cm}
\includegraphics[width=5.0in]{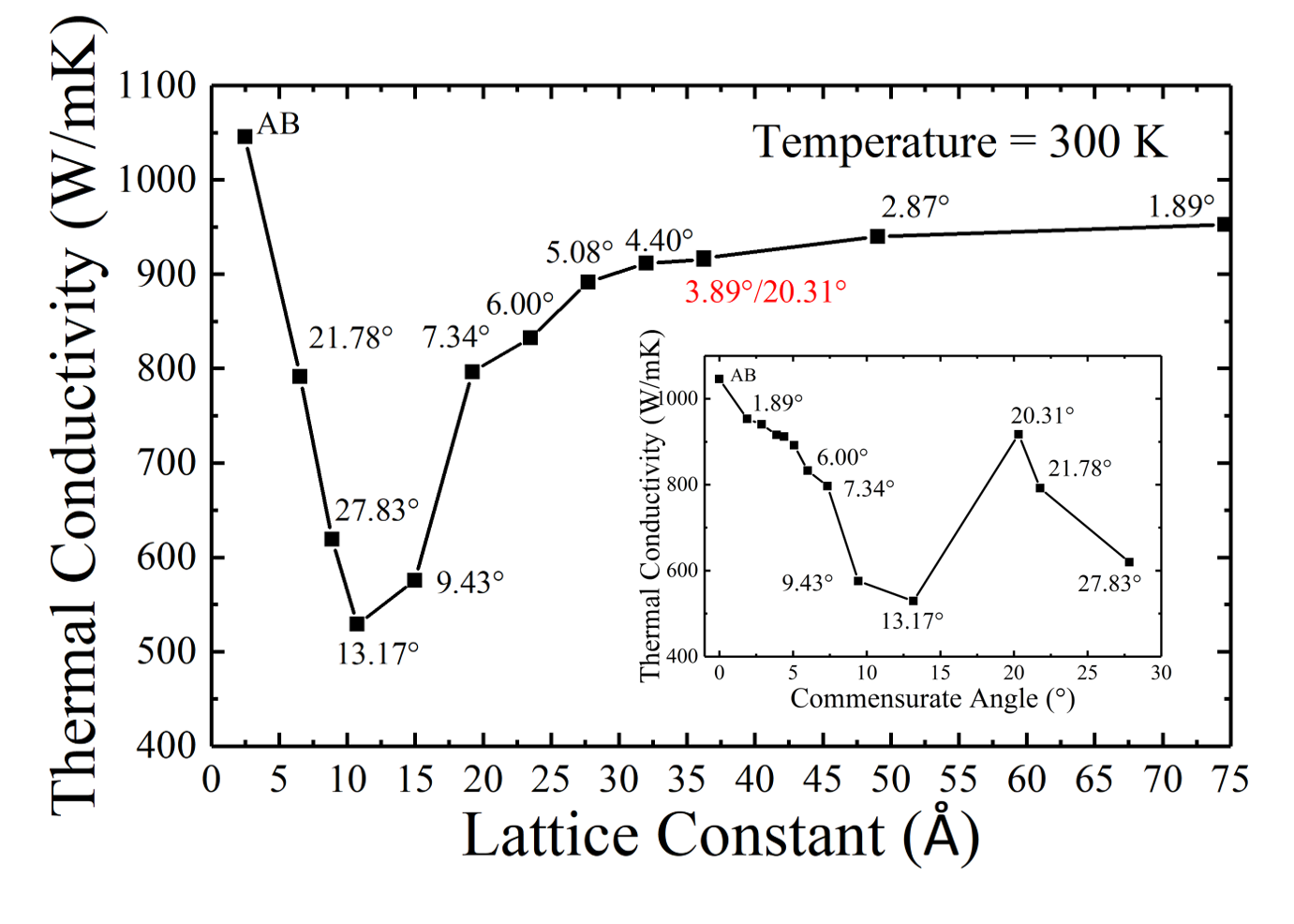} 
\caption{Lattice thermal conductivity of AB-BLG and TBG plotted as a function of 
the primitive commensurate lattice constant. 
The corresponding misorientation angles are shown for each data point.
The inset shows the same data plotted versus twist angle.
}
\label{fig:TC} 
\end{figure}

The calculated room-temperature thermal conductivities for all of the misorientation angles shown in Fig.
\ref{fig:models} are plotted versus their commensurate primitive-cell lattice constants in Fig. \ref{fig:TC}.
The corresponding rotation angles are shown next to each data point.
The first 4 points with the smallest lattice constants have a decreasing linear 
dependence on the commensurate lattice constant, as previously reported \cite{li2018commensurate}.
However, this trend abruptly ends at a commensurate lattice constant of 1.1 nm ($13.17^\circ$), 
where the thermal conductivity reaches a minimum value.
For commensurate lattice constants larger than 1.1 nm, the thermal conductivity monotonically 
increases with increasing lattice 
constant and returns towards the value of the unrotated AB-BLG. 

For the chosen angles below $13^\circ$, the commensurate lattice constants monotonically increase
as the angles decrease.
However, two very different angles, $3.89^\circ$ and $20.31^\circ$ have identical commensurate
lattice constants, and their thermal conductivities are also identical.
This result provides strong evidence that the thermal conductivity of
TBG is a function of the commensurate lattice constant rather than the twist angle. 
To further support that contention, we show the thermal conductivities plotted versus
rotation angle in the inset of Fig. \ref{fig:TC}.  

The calculations in Fig. \ref{fig:TC} were performed with the AIREBO (REBO+LJ) potential.
To verify that the above trend is not an artifact of the interlayer LJ potential, we performed 
a subset of the above calculations using REBO with the
interlayer potential recently developed specifically for twisted multilayer graphene, DRIP.  
The results are shown in Fig. \ref{fig:drip} in the \nameref{sec:Appendix}.
The trends remain the same, with a minimum thermal conductivity occurring at the 
commensurate lattice constant of 1.1 nm.

%
%

From the phonon Boltzmann transport equation \cite{li2014shengbte}, 
$\kappa_l=\frac{1}{N\Omega}\sum_{q,\lambda} \frac{\partial n}{\partial T}(\hbar \omega_\lambda)\nu_\lambda \nu_\lambda \tau_\lambda$, 
two other factors that affect the thermal conductivity are the low-energy phonon frequencies and velocities.
There are 6 low energy phonon branches that originate
from the 3 original acoustic branches, longitudinal (LA), transverse (TA), and out-of-plane (ZA), 
of each individual graphene layer.
We will refer to the 3 acoustic branches that go to 
zero frequency in the BLG and TBG structures as the LA, TA, and ZA modes and the three that have finite frequency
at $\Gamma$ as the LA$_2$, TA$_2$ and ZA$_2$ modes.
We use notation consistent with 
Refs. \cite{singh2011mechanism,2017_Nika_Balandin_RepProgPhys}, 
but we note that the ZA$_2$ mode is often referred to as the ZO' mode in the 
literature describing Raman spectroscopy measurements \cite{Ferrari_Shear_Mode_MLG_NMat12,he2013observation}.
The phonon velocities for the LA and TA phonon 
branches were previously calculated for angles down to {\bf $7.34^\circ$} \cite{cocemasov2013phonons}. 
We now calculate their velocities for angles down to $1.89^\circ$, and we find that over the entire range of angles,
the velocities only vary in the fourth significant digit. 
The velocities of the LA modes lie in the range of
20.03--20.09 km/s, and the velocities of the TA modes lie in the range of 12.83--12.86 km/s.
Thus, the velocities of these two modes play no role in 
explaining the changes in the thermal conductivity with misorientation.

\begin{figure}[H]
\centering
\vspace{0.5cm}
\includegraphics[width=5.0in]{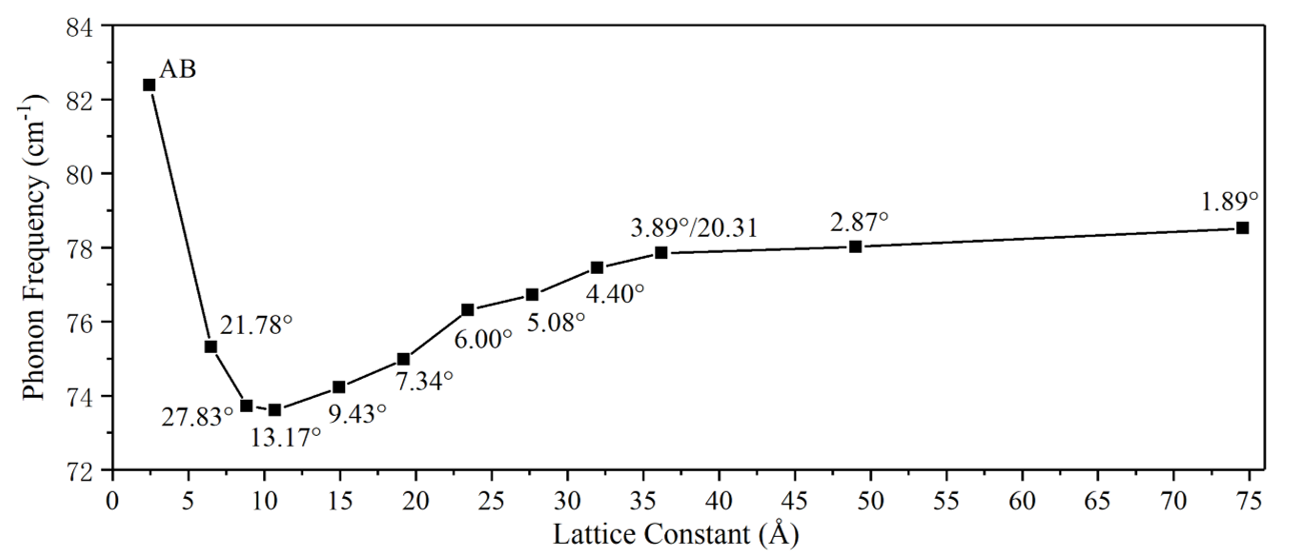} 
\caption{Commensurate lattice constant dependence of the $\Gamma$ point frequency of the ZA$_2$ mode.
}
\label{fig:w_ZA2_v_ZA} 
\end{figure}

The out-of-plane ZA modes in the individual graphene layers strongly couple and split
in frequency when the two layers are brought together to form BLG or TBG.
The $\Gamma$ point frequency of the 
ZA$_2$ mode with AB stacking calculated from LAMMPS is 82.5 cm$^{-1}$.
Fig. \ref{fig:w_ZA2_v_ZA} shows the $\Gamma$ point frequency
of the ZA$_2$ mode, $\w_{\rm ZA_2}$, plotted versus the commensurate lattice constant.
The dependence of the frequency on the commensurate lattice constant
follows the same trend as that of the thermal conductivity.
The mode initially softens, it reaches a minimum frequency at the commensurate lattice constant of 1.1 nm,
and then it begins to harden as the commensurate lattice constant increases.
The ZA$_2$ frequencies for $3.89^\circ$ and $20.31^\circ$ are identical
indicating a dependence on the commensurate lattice constant rather than on the angle.
While the consistency of this trend is interesting, it cannot explain the trends in the
thermal conductivity, since the ZA$_2$ mode is not expected to play a significant role in
thermal transport.

%
%
%

%

\begin{figure}[H]
\centering
\vspace{0.5cm}
\includegraphics[width=5.0in]{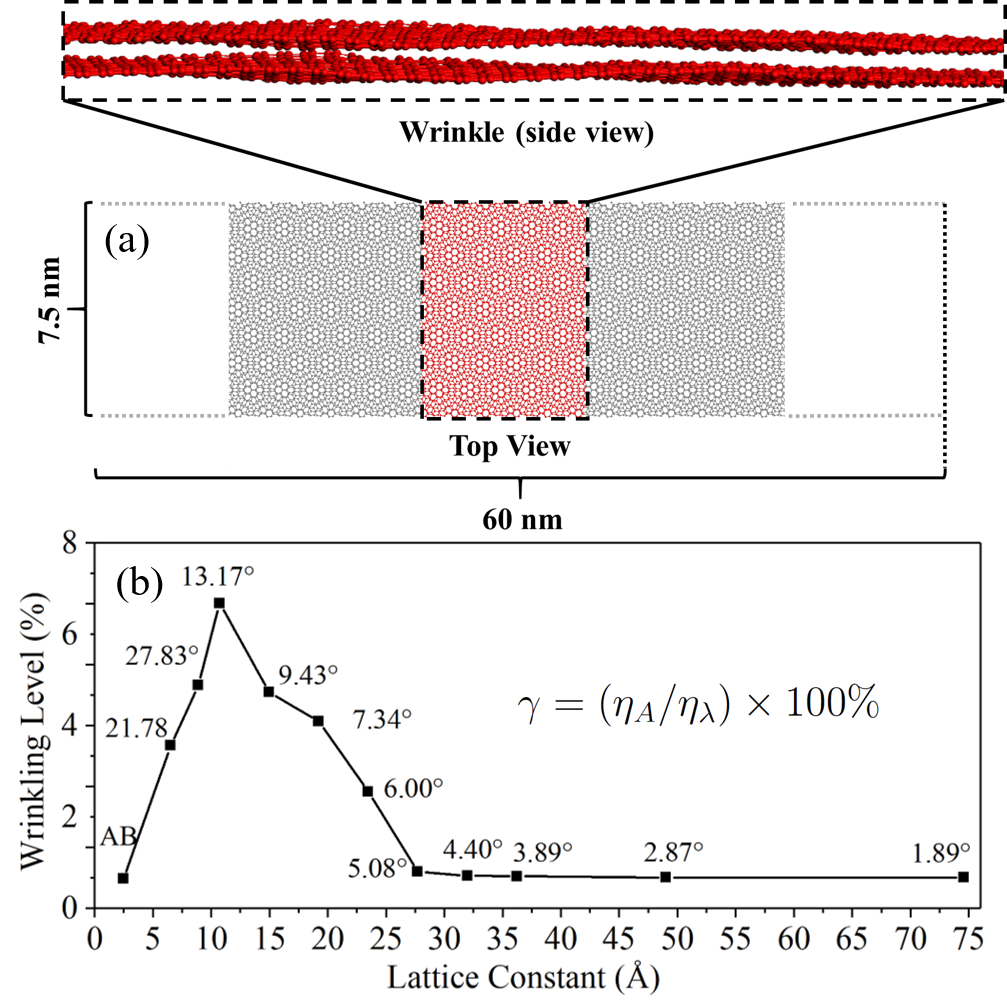} 
\caption{(a) Snapshot of the cross section and top view of the $13.17^\circ$ structure during the 
NEMD simulation. Out-of-plane spatial fluctuations or wrinkling are present.
(b) Wrinkling intensity $\gamma$ as a function of commensurate lattice constant.
}
\label{fig:wrinkle} 
\end{figure}

Since the presence of wrinkles can reduce the thermal conductivity by 
up to 80\% \cite{chen2012thermal,wang2014anisotropic},
and, furthermore, wrinkling will always be present \cite{fasolino2007intrinsic},
we investigate the wrinkling of the TBG structures.
Fig. \ref{fig:wrinkle}(a) shows a snapshot of the $13.17^\circ$ structure during the heat transfer calculation.
Out-of-plane fluctuations or wrinkling are apparent in the cross-sectional view. 
To quantify the intensity of the wrinkling, we plot the unitless metric $\gamma$ (described in Methods)
as a function of the commensurate lattice constant in Fig. \ref{fig:wrinkle}(b).
The wrinkling intensity peaks at the commensurate lattice constant of 1.1 nm corresponding to the minimum
in the thermal conductivity.
The qualitative trends in the wrinkling intensity track those of the thermal conductivity.
The thermal conductivity is lowest when the wrinkling intensity is highest.
%

%
%
%
%
%

\begin{figure}[H]
\centering
\vspace{0.5cm}
\includegraphics[width=5.0in]{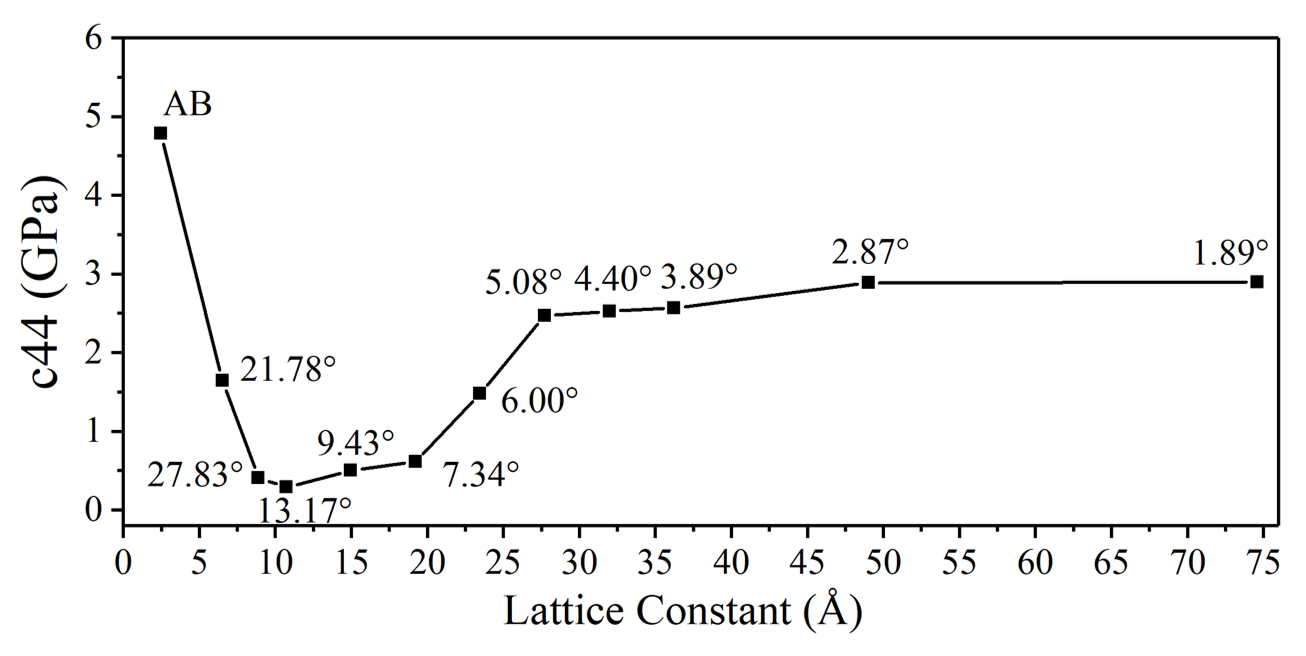} 
\caption{
Elastic constant C$_{44}$ plotted versus the commensurate lattice constant.
}
\label{fig:c44} 
\end{figure}

The ease with which BLG can bend or wrinkle depends on the shear elastic constant 
C$_{44}$ \cite{savini2011bending}.
Therefore, we calculate C$_{44}$ for the structures shown in Fig. \ref{fig:models}, and plot the values
versus commensurate lattice constant in Fig. \ref{fig:c44}.
The trend in C$_{44}$ matches the trends in the thermal conductivity and the wrinkling intensity.
For AB-BLG, C$_{44} = 4.8$ GPa, and this agrees with other experimental and theoretical values
as shown in Table \ref{elastictable}. 
C$_{44}$ reaches a minimum value of 0.293 GPa at the misorientation angle of $13.17^\circ$ with a commensurate
lattice constant of 1.1 nm, and then it returns to 2.9 GPa at the smallest angle of $1.89^\circ$
with a commensurate lattice constant of 7.5 nm.
C$_{44}$ decreases by a factor of 16 between the maximum and minimum value.
At the smallest rotation angle, it is below the AB aligned value by a factor of 1.6.
The calculated numerical values for all angles are given in Table \ref{elastictable}. 
Table \ref{elastictable} also includes calculated values for C$_{11}$,  C$_{12}$, and  C$_{33}$, along with
experimental and theoretical values from other works.
As shown in Table \ref{elastictable}, only C$_{44}$ is affected by interlayer misorientation.

\section{Discussion}
The picture that emerges from the above results is that the interlayer misorientation 
reduces the shear elastic constant C$_{44}$ which increases the wrinkling of the TBG.
The increased out-of-plane wrinkling then reduces the thermal conductivity.
The three parts of this mechanism, reduced C$_{44}$, increased wrinkling, and reduced thermal conductivity,
are consistent with prior results in the literature.
The reduction in C$_{44}$ with misorientation is consistent with previous experimental studies
on Kish graphite \cite{Cij_Graphite_IXS_PRB07} and pyrolitic graphite \cite{Cij_p-graphite_JAP70} 
and a theoretical study of turbostratic graphite \cite{savini2011bending}.
The theoretical study provides a clear description of how a reduction in C$_{44}$ reduces the energy
for out-of-plane wrinkling \cite{savini2011bending}.
Experimental measurements 
found that the average thermal conductivity
of graphene with wrinkles is 27\% lower than that of wrinkle-free graphene \cite{chen2012thermal}. 
NEMD-AIREBO simulations found that a 10\% wrinkling intensity in single layer graphene
resulted in a 20\% decrease in the thermal conductivity 
and a 20\% intensity led to an 80\% decrease \cite{wang2014anisotropic}. 
Therefore, all of the required mechanisms that drive this process are well-established
and validated in the literature.

There have been two previous calculations of the thermal conductivity of 
TBG \cite{minthermal2017,nie2019interlayer}. 
Both studies used LAMMPS with an optimized Tersoff-LJ potential, small structure
sizes (5 nm $\times$ 13 nm) \cite{minthermal2017} (10 nm $\times$ 22 nm) \cite{nie2019interlayer},
incommensurate rotation angles, and open boundaries in the width direction.
The last two items make comparisons with our results problematic. 
Because of the open boundaries, 
the transport was dominated by the edges of the nanoribbons \cite{minthermal2017,nie2019interlayer},
which, because of the incommensurate angles, change as a function of the rotation angle.
In \cite{minthermal2017}, the thermal conductivity decreased as the rotation angle
increased from $0^\circ$ (AB) to $15^\circ$. 
Then, the thermal conductivities of $22.5^\circ$
and $30^\circ$ were larger than that of $0^\circ$ with the maximum occurring at $30^\circ$.
In \cite{nie2019interlayer}, the thermal conductivity monotonically decreased as the rotation angle increased
from $0^\circ$ (AA) to $20^\circ$, and then a local maximum occurred at $30^\circ$. 
The pattern was mirror symmetric as the angle decreased from $60^\circ$ (AB). 
The maximum values occurred for AA and AB stacking, and they were equal.
In both studies, a $30^\circ$ rotation caused one layer to have a zigzag edge and the other layer
to have an armchair edge. 
The relatively smooth edges gave rise to the
maximum (or local maximum) values of the thermal conductivities \cite{minthermal2017,nie2019interlayer}.
However, the presence of angle dependent edge effects prevents meaningful comparisons with our results,
since it is not clear how much of the thermal conductivity 
reduction was due to edge effects and how much was due to other processes.

Previous calculations of C$_{44}$ found a one-order-of-magnitude drop from
4.8 GPa to 0.274 GPa as the commensurate
lattice constant increased from 2.46 {\AA} (AB) to 6.51 {\AA} ($21.78^\circ$) \cite{savini2011bending}.
As the lattice constant increased further, 
C$_{44}$ gradually declined to a minimum average 
value of 0.2 GPa at a commensurate lattice constant of 2.56 nm corresponding to a rotation
angle of $11.0^\circ$.
The calculations were performed using density functional theory (DFT). 

Prior calculations of the ZA$_2$ frequency found a drop from 95 cm$^{-1}$ to 89.5 cm$^{-1}$
as the commensurate lattice constant increased from 2.46 {\AA} (AB) to 
6.51 {\AA} ($21.78^\circ$) \cite{cocemasov2013phonons}.
After the initial decrease, there was a slight monotonic decline to  89.1 cm$^{-1}$
as the commensurate lattice constant was increased to 1.9 nm 
corresponding to a misorientation angle of $7.34^\circ$.
These calculations used the Born-von Karman (BvK) model for the intralayer forces and the LJ potential
for interlayer forces.

One significant difference between our NEMD and MD simulations and the BvK model or DFT
calculations is that our simulations explicitly take into account finite temperature effects
and time-dependent thermal fluctuations.
DFT is a zero-temperature theory. 
In the BvK approach,
there is no relaxation of the structure so that the geometry of the layers remains ideally flat.
In our NEMD and MD calculations, the effects of finite temperature
and out-of-plane wrinkling
are included both in the thermal conductivity calculations
and in the construction of the dynamical matrix for the calculations of the phonon
spectra.

What is unique to our results is the prediction of non-monotonic behavior of the thermal conductivity 
with respect to the commensurate lattice constant.
For the small angle rotations, the commensurate lattice constants become
extremely large.
Our calculations of the thermal conductivity, C$_{44}$, and $\w_{\rm ZA_2}$ all show
a return to a value similar to, but less than the value of the aligned AB structure
as the twist angle is reduced to $1.89^\circ$.
If we extrapolate the trends in C$_{44}$, and $\w_{\rm ZA_2}$ observed 
previously \cite{savini2011bending,cocemasov2013phonons},
the values would continuously decline as the twist angle approached $0^\circ$, followed by a sudden
large discontinuity as the angle became exactly $0^\circ$.
At small twist angles $\theta \lesssim 1^\circ$, there are large regions that are close to AA stacking,
large regions that are close to AB stacking, and connecting regions that are misaligned.
Whether it is appropriate to view the thermal conductivity of such structures 
as an average of different macroscopic regions of aligned structures 
and misaligned structures is unclear, but
such a view would be consistent with the small-angle trend that we observe.
If such a perspective is correct, it raises the question of what length scale determines when such
a view is permissible or not.
%
%
%
%

\section{Summary, Conclusions, and Open Questions}
Large scale room temperature NEMD calculations of the thermal conductivity of twisted bilayer graphene
find a non-monotonic dependence of the thermal conductivity on the commensurate
lattice constant. 
At a commensurate lattice constant of 1.1 nm corresponding
to an angle of $13.2^\circ$, 
the thermal conductivity falls to 50\% of the value of the aligned AB structure.
As the commensurate lattice constant increases,
the thermal conductivity also increases and reaches 91\% of the AB value
at a commensurate lattice constant of 7.5 nm corresponding to an angle of $1.89^\circ$.
The commensurate-lattice-constant-dependent trends in the thermal 
conductivity are also followed by the trends in the shear elastic constant C$_{44}$, the wrinkling intensity,
and the frequency of the out-of-plane ZA$_2$ mode.
The picture that emerges from these results is that the interlayer misorientation 
reduces the shear elastic constant C$_{44}$, 
the reduced shear elastic constant allows increased wrinkling of the TBG, and
the increased wrinkling reduces the thermal conductivity.
The small-angle approach of the thermal conductivity towards its value in the aligned structure
raises the question of how response functions approach their aligned values
as the twist angle approaches $0^\circ$.
Is the approach gradual, discontinuous, or a combination of the two?
%
\section*{Acknowledgments}

This work was supported in part by the National Science Foundation under awards 1307671 and 1433395.
Numerical simulations were supported in part by  the Spins and Heat in Nanoscale Electronic Systems (SHINES) an Energy Frontier Research Center funded by the U.S. Department of Energy, Office of Science, Basic Energy Sciences under Award {\#}DE-SC0012670.
This work used the Extreme Science and Engineering Discovery
Environment (XSEDE) \cite{towns2014xsede}
which is supported by National
Science Foundation Grant No. ACI-1548562 and allocation
ID TG-DMR130081. 
Used resources include Stampede and
Comet.
%
\section*{Appendix}
\label{sec:Appendix}
Table. \ref{elastictable} shows the calculated elastic constants for each rotation angle along with experimental
values and values calculated from DFT.
Only C$_{44}$ is affected by misorientation.
\begin{table}[H]
\centering
\caption{\label{elastictable}
Calculated elastic constants of AB-BLG and TBG. from prior DFT calculation and from our MD calculation using the hybrid REBO and LJ potentials}
\footnotesize
\begin{tabular}{@{}lllll}
\br
Rotation angle ($^{\circ}$ )&C$_{11}$ (GPa)&C$_{12}$ (GPa)&C$_{33}$ (GPa)&C$_{44}$ (GPa)\\
\mr
0(EXP)\cite{Cij_Graphite_IXS_PRB07}  & 1109$\pm$16 & 139$\pm$36 & 38.7$\pm$7 & 5$\pm$3\\
0(DFT)\cite{savini2011bending}  & 1109 & 175 & 42 & 4.8\\
0 & 1023.6 & 227.1 & 42.3 & 4.79\\
21.78 & 1023.6 & 227.2 & 42.6 & 1.65\\
27.83 & 1023.8 & 227.3 & 42.7 & 0.411\\
13.17 & 1023.3 & 227.8 & 42.7 & 0.293\\
9.43 & 1023.7 & 227.4 & 42.7 & 0.503\\
7.34 & 1023.7 & 227.4 & 42.6 & 0.618\\
6.00 & 1023.6 & 227.5 & 42.7 & 1.485\\
5.08 & 1023.8 & 227.6 & 42.6 & 2.47\\
4.40 & 1023.5 & 227.6 & 42.6 & 2.53\\
3.89 & 1023.2 & 227.9 & 42.6 & 2.57\\
2.87 & 1023.5 & 227.9 & 42.6 & 2.89\\
1.89 & 1023.4 & 227.9 & 42.6 & 2.90\\
\br
\end{tabular}\\
\end{table}
\normalsize

\begin{figure}[H]
\centering
\vspace{0.5cm}
\includegraphics[width=5.0in]{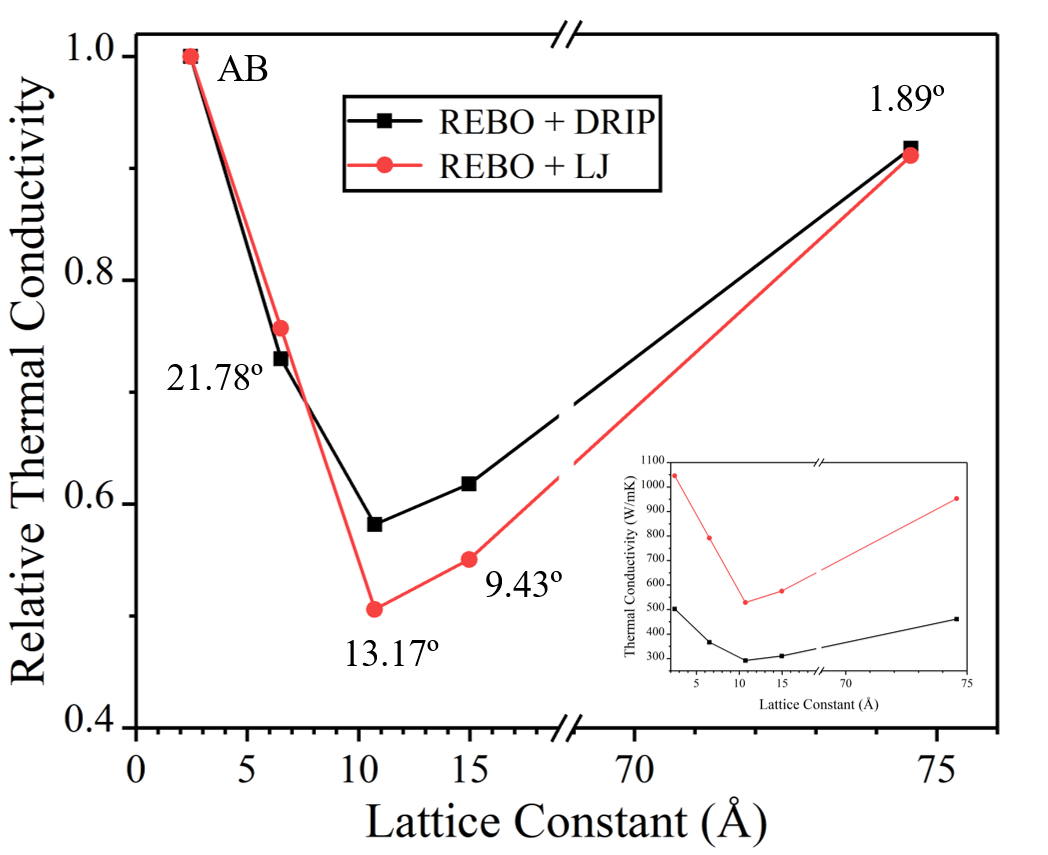} 
\caption{
Lattice constant dependent relative thermal conductivity (normalized to the AB value) 
for two different interlayer potentials as shown in the legend.
The inset shows the absolute values. 
The REBO+LJ values are the extracted $L=\infty$ values from Fig. \ref{fig:TC}. 
The REBO+DRIP values are from the finite length $130$ nm structures.
Thus, these values are expected to be quantitatively lower.
}
\label{fig:drip}
\end{figure}

To verify that the trends shown in Fig. \ref{fig:TC} are not an artifact of the AIREBO implementation
of the LJ potential, we performed a subset of the calculations using REBO+DRIP. 
Five commensurate angles are selected: 0$^\circ$ (AB), 21.78$^\circ$, 13.17$^\circ$, 9.43$^\circ$ and 1.89$^\circ$.
Instead of running multiple simulations of different lengths for each angle and extracting the $L=\infty$ value
of the thermal conductivity, we choose one length of $\sim 130$ nm for each angle and a width of $\sim 15$ nm.
Since the structures are composed of integer numbers of primitive cells that have different sizes, 
the actual widths lie between 149.12 {\AA} to 150.09 \AA, and the lengths range from 1284.88 {\AA} to 1304.14 \AA. 
Due to the finite lengths, the quantitative values will be lower, than those in Fig. \ref{fig:TC}, however,
here, we only wish to confirm the non-monotonic trend of the thermal conductivity with the commensurate lattice constant.
All other settings related to the NEMD simulations, periodic boundary conditions in the width direction, 
and temperature (300 K) are as described in the \nameref{sec:Methods} section. 
It is clear from the results shown in Fig. \ref{fig:drip} that the trends in the thermal conductivity with respect
to the commensurate lattice constant are unaffected by the choice of the interlayer potential.

\newpage
\section*{References}
\bibliographystyle{unsrt}
\bibliography{ref_RL_CL}

\begin{thebibliography}{10}

\bibitem{mele2010commensuration}
E.~J. Mele.
\newblock Commensuration and interlayer coherence in twisted bilayer graphene.
\newblock {\em Physical Review B}, 81(16):161405, 2010.

\bibitem{shallcross2010electronic}
S.~Shallcross, S.~Sharma, E.~Kandelaki, and O.~A. Pankratov.
\newblock Electronic structure of turbostratic graphene.
\newblock {\em Physical Review B}, 81(16):165105, 2010.

\bibitem{Flat_Bands_TGB_Morell_PRB10}
E.~Su\'arez~Morell, J.~D. Correa, P.~Vargas, M.~Pacheco, and Z.~Barticevic.
\newblock Flat bands in slightly twisted bilayer graphene: Tight-binding
  calculations.
\newblock {\em Phys. Rev. B}, 82:121407, Sep 2010.

\bibitem{MacDonald_PNAS11}
R.~Bistritzer and A.~H. MacDonald.
\newblock Moir{\'e} bands in twisted double-layer graphene.
\newblock {\em Proceedings of the National Academy of Sciences},
  108(30):12233--12237, 2011.

\bibitem{bistritzer2010transport}
R.~Bistritzer and A.~H. MacDonald.
\newblock Transport between twisted graphene layers.
\newblock {\em Physical Review B}, 81(24):245412, 2010.

\bibitem{Habib_Graphite_APL13}
K.~M.~M. Habib, S.~S. Sylvia, S.~Ge, M.~Neupane, and R.~K. Lake.
\newblock The coherent interlayer resistance of a single, rotated interface
  between two stacks of ab graphite.
\newblock {\em Appl. Phys. Lett.}, 103(24):243114, 2013.

\bibitem{koren2016coherent}
E.~Koren, I.~Leven, E.~L{\"o}rtscher, A.~Knoll, O.~Hod, and U.~Duerig.
\newblock Coherent commensurate electronic states at the interface between
  misoriented graphene layers.
\newblock {\em Nature nanotechnology}, 11(9):752, 2016.

\bibitem{Shepard_NL16}
T.~Chari, R.~Ribeiro-Palau, C.~R. Dean, and K.~Shepard.
\newblock Resistivity of rotated graphite–graphene contacts.
\newblock {\em Nano Letters}, 16(7):4477--4482, 2016.

\bibitem{Bernevig_All_Angles_Topo_PRL19}
Z.~Song, Z.~Wang, W.~Shi, G.~Li, C.~Fang, and B.~A. Bernevig.
\newblock All magic angles in twisted bilayer graphene are topological.
\newblock {\em Physical review letters}, 123(3):036401, 2019.

\bibitem{cao2018unconventional}
Y.~Cao, V.~Fatemi, S.~Fang, K.~Watanabe, T.~Taniguchi, E.~Kaxiras, and
  P.~Jarillo-Herrero.
\newblock Unconventional superconductivity in magic-angle graphene
  superlattices.
\newblock {\em Nature}, 556(7699):43, 2018.

\bibitem{Tuning_SC_TBG_CDean_Sc19}
M.~Yankowitz, S.~Chen, H.~Polshyn, Y.~Zhang, K.~Watanabe, T.~Taniguchi,
  D.~Graf, A.~F. Young, and Cory~R. Dean.
\newblock Tuning superconductivity in twisted bilayer graphene.
\newblock {\em Science}, 363(6431):1059--1064, 2019.

\bibitem{carozo2011raman}
V.~Carozo, C.~M. Almeida, E.~H.~M. Ferreira, L.~G. Cancado, C.~A. Achete, and
  A.~Jorio.
\newblock Raman signature of graphene superlattices.
\newblock {\em Nano letters}, 11(11):4527--4534, 2011.

\bibitem{havener2012angle}
R.~W. Havener, H.~Zhuang, L.~Brown, R.~G. Hennig, and J.~Park.
\newblock Angle-resolved raman imaging of interlayer rotations and interactions
  in twisted bilayer graphene.
\newblock {\em Nano letters}, 12(6):3162--3167, 2012.

\bibitem{kim2012raman}
K.~Kim, S.~Coh, L.~Z. Tan, W.~Regan, J.~M. Yuk, E.~Chatterjee, M.~F. Crommie,
  M.~L. Cohen, S.~G. Louie, and A.~Zettl.
\newblock Raman spectroscopy study of rotated double-layer graphene:
  misorientation-angle dependence of electronic structure.
\newblock {\em Physical review letters}, 108(24):246103, 2012.

\bibitem{he2013observation}
R.~He, T.~F. Chung, C.~Delaney, C.~Keiser, L.~A. Jauregui, P.~M. Shand, C.~C.
  Chancey, Y.~Wang, J.~Bao, and Y.~P. Chen.
\newblock Observation of low energy raman modes in twisted bilayer graphene.
\newblock {\em Nano letters}, 13(8):3594--3601, 2013.

\bibitem{campos2013raman}
J.~Campos-Delgado, L.~G. Can{\c{c}}ado, C.~A. Achete, A.~Jorio, and J.~P.
  Raskin.
\newblock Raman scattering study of the phonon dispersion in twisted bilayer
  graphene.
\newblock {\em Nano Research}, 6(4):269--274, 2013.

\bibitem{lui2012observation}
C.~H. Lui, L.~M. Malard, S.~H. Kim, G.~Lantz, F.~E. Laverge, R.~Saito, and
  T.~F. Heinz.
\newblock Observation of layer-breathing mode vibrations in few-layer graphene
  through combination raman scattering.
\newblock {\em Nano letters}, 12(11):5539--5544, 2012.

\bibitem{coh2013theory}
S.~Coh, L.~Z. Tan, S.~G. Louie, and M.~L. Cohen.
\newblock Theory of the raman spectrum of rotated double-layer graphene.
\newblock {\em Physical Review B}, 88(16):165431, 2013.

\bibitem{righi2013resonance}
A.~Righi, P.~Venezuela, H.~Chacham, S.~D. Costa, C.~Fantini, R.~S. Ruoff,
  L.~Colombo, W.~S. Bacsa, and M.~A. Pimenta.
\newblock Resonance raman spectroscopy in twisted bilayer graphene.
\newblock {\em Solid State Communications}, 175:13--17, 2013.

\bibitem{ramnani2017raman}
P.~Ramnani, M.~R. Neupane, S.~Ge, A.~A. Balandin, R.~K. Lake, and
  A.~Mulchandani.
\newblock Raman spectra of twisted cvd bilayer graphene.
\newblock {\em Carbon}, 123:302--306, 2017.

\bibitem{eliel2018intralayer}
G.~S.~N. Eliel, M.~V.~O. Moutinho, A.~C. Gadelha, A.~Righi, L.~C. Campos, H.~B.
  Ribeiro, P.~W. Chiu, K.~Watanabe, T.~Taniguchi, P.~Puech, et~al.
\newblock Intralayer and interlayer electron--phonon interactions in twisted
  graphene heterostructures.
\newblock {\em Nature communications}, 9(1):1221, 2018.

\bibitem{cocemasov2013phonons}
A.~I. Cocemasov, D.~L. Nika, and A.~A. Balandin.
\newblock Phonons in twisted bilayer graphene.
\newblock {\em Physical Review B}, 88(3):035428, 2013.

\bibitem{cocemasov2015engineering}
A.~I. Cocemasov, D.~L. Nika, and A.~A. Balandin.
\newblock Engineering of the thermodynamic properties of bilayer graphene by
  atomic plane rotations: the role of the out-of-plane phonons.
\newblock {\em Nanoscale}, 7(30):12851--12859, 2015.

\bibitem{nika2014specific}
D.~L. Nika, A.~I. Cocemasov, and A.~A. Balandin.
\newblock Specific heat of twisted bilayer graphene: Engineering phonons by
  atomic plane rotations.
\newblock {\em Applied Physics Letters}, 105(3):031904, 2014.

\bibitem{li2018commensurate}
C.~Li, B.~Debnath, X.~Tan, S.~Su, K.~Xu, S.~Ge, M.~R. Neupane, and R.~K. Lake.
\newblock Commensurate lattice constant dependent thermal conductivity of
  misoriented bilayer graphene.
\newblock {\em Carbon}, 138:451--457, 2018.

\bibitem{Bernevig_Phonon_Driven_TBG_Superconductor}
B.~Lian, Z.~Wang, and B.~A. Bernevig.
\newblock Twisted bilayer graphene: A phonon-driven superconductor.
\newblock {\em Phys. Rev. Lett.}, 122:257002, Jun 2019.

\bibitem{li2014thermal}
Hongyang Li, Hao Ying, Xiangping Chen, Denis~L Nika, Alexandr~I Cocemasov,
  Weiwei Cai, Alexander~A Balandin, and Shanshan Chen.
\newblock Thermal conductivity of twisted bilayer graphene.
\newblock {\em Nanoscale}, 6(22):13402--13408, 2014.

\bibitem{mao2016atomistic}
R.~Mao, Y.~Chen, and K.~W. Kim.
\newblock Atomistic modeling of phonon transport in turbostratic graphitic
  structures.
\newblock {\em Journal of Applied Physics}, 119(20):204305, 2016.

\bibitem{minthermal2017}
M.~H. Wang, Y.~E. Xie, and Y.~P. Chen.
\newblock Thermal transport in twisted few-layer graphene.
\newblock {\em Chinese Physics B}, 26(11):116503, 2017.

\bibitem{nie2019interlayer}
X.~Nie, L.~Zhao, S.~Deng, Y.~Zhang, and Z.~Du.
\newblock How interlayer twist angles affect in-plane and cross-plane thermal
  conduction of multilayer graphene: A non-equilibrium molecular dynamics
  study.
\newblock {\em International Journal of Heat and Mass Transfer}, 137:161--173,
  2019.

\bibitem{ghosh2008extremely}
S.~Ghosh, I.~Calizo, D.~Teweldebrhan, E.~P. Pokatilov, D.~L. Nika, A.~A
  Balandin, W.~Bao, F.~Miao, and C.~N. Lau.
\newblock Extremely high thermal conductivity of graphene: Prospects for
  thermal management applications in nanoelectronic circuits.
\newblock {\em Applied Physics Letters}, 92(15):151911, 2008.

\bibitem{muller1997simple}
F.~M{\"u}ller-Plathe.
\newblock A simple nonequilibrium molecular dynamics method for calculating the
  thermal conductivity.
\newblock {\em The Journal of chemical physics}, 106(14):6082--6085, 1997.

\bibitem{plimpton2007lammps}
S.~Plimpton, P.~Crozier, and A.~Thompson.
\newblock Lammps-large-scale atomic/molecular massively parallel simulator.
\newblock {\em Sandia National Laboratories}, 18:43--43, 2007.

\bibitem{koukaras2015phonon}
E.~N. Koukaras, G.~Kalosakas, C.~Galiotis, and K.~Papagelis.
\newblock Phonon properties of graphene derived from molecular dynamics
  simulations.
\newblock {\em Scientific reports}, 5:12923, 2015.

\bibitem{tersoff1989modeling}
J.~Tersoff.
\newblock Modeling solid-state chemistry: Interatomic potentials for
  multicomponent systems.
\newblock {\em Physical Review B}, 39(8):5566, 1989.

\bibitem{lindsay2010optimized}
L.~Lindsay and D.~A. Broido.
\newblock Optimized tersoff and brenner empirical potential parameters for
  lattice dynamics and phonon thermal transport in carbon nanotubes and
  graphene.
\newblock {\em Physical Review B}, 81(20):205441, 2010.

\bibitem{Brenner_PRB90}
D.~W. Brenner.
\newblock Empirical potential for hydrocarbons for use in simulating the
  chemical vapor deposition of diamond films.
\newblock {\em Physical review B}, 42(15):9458, 1990.

\bibitem{stuart2000reactive}
S.~J. Stuart, A.~B. Tutein, and J.~A. Harrison.
\newblock A reactive potential for hydrocarbons with intermolecular
  interactions.
\newblock {\em The Journal of chemical physics}, 112(14):6472--6486, 2000.

\bibitem{brenner2002second}
D.~W. Brenner, O.~A. Shenderova, J.~A. Harrison, S.~J. Stuart, B.~Ni, and S.~B.
  Sinnott.
\newblock A second-generation reactive empirical bond order (rebo) potential
  energy expression for hydrocarbons.
\newblock {\em Journal of Physics: Condensed Matter}, 14(4):783, 2002.

\bibitem{los2003intrinsic}
J.~H. Los and A.~Fasolino.
\newblock Intrinsic long-range bond-order potential for carbon: Performance in
  monte carlo simulations of graphitization.
\newblock {\em Physical Review B}, 68(2):024107, 2003.

\bibitem{abell1985empirical}
G.~C. Abell.
\newblock Empirical chemical pseudopotential theory of molecular and metallic
  bonding.
\newblock {\em Physical Review B}, 31(10):6184, 1985.

\bibitem{wen2018dihedral}
M.~Wen, S.~Carr, S.~Fang, E.~Kaxiras, and Ellad~B. Tadmor.
\newblock Dihedral-angle-corrected registry-dependent interlayer potential for
  multilayer graphene structures.
\newblock {\em Physical Review B}, 98(23):235404, 2018.

\bibitem{huang2006evaluation}
Z.~X. Huang and Z.~A. Tang.
\newblock Evaluation of momentum conservation influence in non-equilibrium
  molecular dynamics methods to compute thermal conductivity.
\newblock {\em Physica B: Condensed Matter}, 373(2):291--296, 2006.

\bibitem{kong2011phonon}
L.~T. Kong.
\newblock Phonon dispersion measured directly from molecular dynamics
  simulations.
\newblock {\em Computer Physics Communications}, 182(10):2201--2207, 2011.

\bibitem{wang2014anisotropic}
C.~Wang, Y.~Liu, L.~Li, and H.~Tan.
\newblock Anisotropic thermal conductivity of graphene wrinkles.
\newblock {\em Nanoscale}, 6(11):5703--5707, 2014.

\bibitem{li2014shengbte}
W.~Li, J.~Carrete, N.~A. Katcho, and N.~Mingo.
\newblock Shengbte: A solver of the boltzmann transport equation for phonons.
\newblock {\em Computer Physics Communications}, 185(6):1747--1758, 2014.

\bibitem{singh2011mechanism}
D.~Singh, J.~Y. Murthy, and T.~S. Fisher.
\newblock Mechanism of thermal conductivity reduction in few-layer graphene.
\newblock {\em Journal of Applied Physics}, 110(4):044317, 2011.

\bibitem{2017_Nika_Balandin_RepProgPhys}
L.~N. Denis and A.~B. Alexander.
\newblock Phonons and thermal transport in graphene and graphene-based
  materials.
\newblock {\em Reports on Progress in Physics}, 80(3):036502, 2017.

\bibitem{Ferrari_Shear_Mode_MLG_NMat12}
P.~H. Tan, W.~P. Han, W.~J. Zhao, Z.~H. Wu, K.~Chang, H.~Wang, Y.~F. Wang,
  N.~Bonini, N.~Marzari, N.~Pugno, G.~Savini, A.~Lombardo, and A.~C. Ferrari.
\newblock The shear mode of multilayer graphene.
\newblock {\em Nature Materials}, 11(4):294--300, 2012.

\bibitem{chen2012thermal}
S.~Chen, Q.~Li, Q.~Zhang, Y.~Qu, H.~Ji, R.~S. Ruoff, and W.~Cai.
\newblock Thermal conductivity measurements of suspended graphene with and
  without wrinkles by micro-raman mapping.
\newblock {\em Nanotechnology}, 23(36):365701, 2012.

\bibitem{fasolino2007intrinsic}
A.~Fasolino, J.~H. Los, and M.~I. Katsnelson.
\newblock Intrinsic ripples in graphene.
\newblock {\em Nature materials}, 6(11):858, 2007.

\bibitem{savini2011bending}
G.~Savini, Y.~J. Dappe, S.~{\"O}berg, J.~C. Charlier, M.~I. Katsnelson, and
  A.~Fasolino.
\newblock Bending modes, elastic constants and mechanical stability of
  graphitic systems.
\newblock {\em Carbon}, 49(1):62--69, 2011.

\bibitem{Cij_Graphite_IXS_PRB07}
A.~Bosak, M.~Krisch, M.~Mohr, J.~Maultzsch, and C.~Thomsen.
\newblock Elasticity of single-crystalline graphite: Inelastic x-ray scattering
  study.
\newblock {\em Phys. Rev. B}, 75:153408, Apr 2007.

\bibitem{Cij_p-graphite_JAP70}
O.~L. Blakslee, D.~G. Proctor, E.~J. Seldin, G.~B. Spence, and T.~Weng.
\newblock Elastic constants of compression{-}annealed pyrolytic graphite.
\newblock {\em Journal of Applied Physics}, 41(8):3373--3382, 1970.

\bibitem{towns2014xsede}
J.~Towns, T.~Cockerill, M.~Dahan, I.~Foster, K.~Gaither, A.~Grimshaw,
  V.~Hazlewood, S.~Lathrop, D.~Lifka, G.~D. Peterson, et~al.
\newblock Xsede: accelerating scientific discovery.
\newblock {\em Computing in Science \& Engineering}, 16(5):62--74, 2014.

\end{thebibliography}

\end{document}